\documentclass[doublecol]{epl2} 

\title{Network theory approach for data evaluation in the dynamic force spectroscopy of biomolecular interactions }
\shorttitle{Network approach for evaluation of single-molecule force spectroscopy signals} 

\author{Jelena \v Zivkovi\'c\inst{1} Marija Mitrovi\'c \inst{3}  Luuk Janssen \inst{2}  Hans A. Heus \inst{2}    Bosiljka Tadi\'c \inst{3} Sylvia Speller \inst{1}}
\shortauthor{J. \v Zivkovi\'c \etal}

\institute{                    
  \inst{1} Scanning Probe Microscopy Group, Insitute for Molecules and Materials,  Radboud University, Nijmegen, Netherlands
\inst{2}Department of Biophysical Chemistry, Insitute for Molecules and Materials,
Radboud University , Nijmegen, Netherlands
  \inst{3} Department of theoretical physics, Jo\v zef Stefan Institute, Ljubljana, Slovenia
}

\abstract{Investigations of molecular bonds between single molecules and molecular complexes by the dynamic force spectroscopy are subject to large fluctuations at nanoscale and possible other aspecific binding, which mask the experimental output.  Big efforts are devoted to develop  methods for effective selection of the relevant experimental data,  before taking the quantitative analysis of bond parameters. Here we present a methodology which is based on the application of graph theory.
The force--distance curves corresponding to  repeated pulling events are mapped onto their correlation network (mathematical graph). On these graphs the groups of similar curves appear as topological modules, which are identified using the spectral analysis of  graphs. 
We demonstrate the approach  by analyzing a large ensemble of the force--distance curves measured on: ssDNA--ssDNA, peptide--RNA (system
from HIV1), and peptide--Au surface.
Within our data sets the methodology systematically separates subgroups of curves which are related to different intermolecular interactions and to spatial arrangements  in which the molecules are brought together and/or pulling speeds. This demonstrates the sensitivity of the method to the spatial degrees of freedom, suggesting potential applications in the case of large molecular complexes  and situations with multiple binding sites. 
}

\begin{document}

\maketitle

\section{Introduction}It has been recognized  recently \cite{barkai2008} that the signals generated at a single molecule (or another nano-size object) differ from signals obtained in large-scale systems consisting of ensemble of molecules. In particular, enhanced fluctuations, randomness and irreproducibility of the signals are observed in single-molecule measurements. A representative example is the   mechanical signal generated in the dynamic force spectroscopy (DFS) of molecular bonds \cite{evansritchie1997,strunz1999}. The force spectroscopy  
 of single molecules and molecular complexes  has become a leading methodology for measuring biomolecular unbinding forces, which form the bases of biologically relevant molecular processes \cite{DSFreview2000}. For instance, some recently studied examples include
measurements of fundamental biomolecular forces in DNA unzipping \cite{koch2003}, ALCAM-ALCAM \cite{riet2007}, 
peptide- antibody \cite{sulchek2005},  RNA- protein \cite{RNA2009} interactions, etc.

In a typical pulling experiment in DFS based on the Atomic Force Microscopy, the ligand and receptor molecule are attached via polymer linkers on the AFM tip and the solid  support (e.g
glass, mica, gold- surface). The molecules are brought close to each other for certain contact time allowing them to form a bond and then pulled away until the bond breaks. The process is repeated many times. In each pulling event, changes in the 
deflection of the AFM  cantilever as a function of distance are measured.
Knowing the spring constant of the cantilever, this data can be calculated into
distance dependent forces, resulting in so called force-distance curves.
From these curves different parameters can be obtained, such as force needed to
break a certain bond and the force loading rate. Further quantitative analysis of these data sets requires an elaborated theoretical framework \cite{evansritchie1997} in order to extract the parameters of binding potential, the bond strength and the survival time. 
The applied force reduces the barrier between the bound and dissociated states of the binding molecules, allowing to estimate the force at which the barrier disappears and to measure the dissociation rate $\lambda(F)$. Then the {\it natural} dissociation rate at vanishing force is extracted using an appropriate theoretical framework. 
Since the pioneering work \cite{evansritchie1997}, assuming the exponential dependence of the dissociation rate on the applied force, the theory of the dynamic force spectroscopy evolved in two major directions: (i) incorporating a phenomenological distribution of the bond parameters \cite{raible2006}, and (ii) assuming a specific type of the binding potential, from which the dissociation rate $\lambda (F)$ is computed exactly \cite{dudko2006,dudko2008}.   

The pulling--disruption process of the molecular bonds, for instance in constant-velocity experiments used in this work,  is both nonlinear and stochastic. Nonlinearity, which  is manifested in force-dependent loading rate $\dot{F}(F)$, can be related to the internal degrees of freedom of the (large) molecules and the spacers. 
Fluctuations in the disruption force and loading rate originate from the diffusion of binding molecules in solvent, whose effects depend on the binding potential and pulling speed. 
Thus the bond parameters can be determined in terms of  probability, using the appropriate theoretical concept. 
In addition to the molecular bond of interest (specific binding), other non-specific processes may occur at the same contact time and range of forces: mis-binding or binding at different site, or unspecific interaction of the molecule with linkers  and surface. 
It is, therefore, of great importance to be able to select the force-distance curves which are related to true binding {\it prior} to the quantitative analysis of the bond parameters.  It is a general believe (see also discussion in \cite{fuhrmann2008}) that, apart from the fluctuations, {\it the force-distance curves originating from the same binding process share strong similarity.} Given a large amount of data in a particular experiment, this is a technically demanding task, which requires automated computational approach.  Recently two methods were proposed based on pattern-recognition \cite{marsico2006} and master-curve fitting \cite{fuhrmann2008}.

In this work  we present a new  approach for systematic selection of  groups of mutually similar force--distance curves and demonstrate it on sets of data from peptide--RNA complex from HIV1. 
Our methodology is based on the theory of complex networks and their spectral analysis \cite{boccaletti,donetti2004,mitrovic2008b}.  Mapping multichannel datasets onto a network representation has proved as a useful tool  \cite{benjacob2006} in the analysis of many complex dynamical systems, for example stock-market time series \cite{mantegna1999}, gene-expression signals \cite{zivkovic2006,zivkovic2009}, neural activity signals \cite{zemanova2000,eshel2004,book_kurths2008}, climat phase fluctuations \cite{shlomo2009}, and  information traffic time series \cite{tadic2009}.  Compared to these examples, where the time-series are measured at each unit of an extended interacting system, our approach here is to map the  force--distance signal measured in repeated experiments on the  same molecular complex  with many degrees of freedom.  
To demonstrate our approach, we analyze a large dataset of force--distance curves measured under different conditions in  RNA--peptide complex (HIV1-Rev peptide binding at the high affinity site in the stem-loop of the RRE on viral RNA) \cite{Jelena-experimental2009} and, for comparison, a set of curves for ssDNA--ssDNA binding and two control sets.  The filtered correlation matrix mapped onto a binary graph exhibits modular structure, with the subgraphs of nodes (curve index) grouped according to their similarity over the entire data sets. We then look for the  force-curves in each of the subgraphs and find the underlying reasons for their clustering.

\section{ Description of the Experimental Data}
\begin{figure}
\centering
\begin{tabular}{cc}
\resizebox{18pc}{!}{\includegraphics{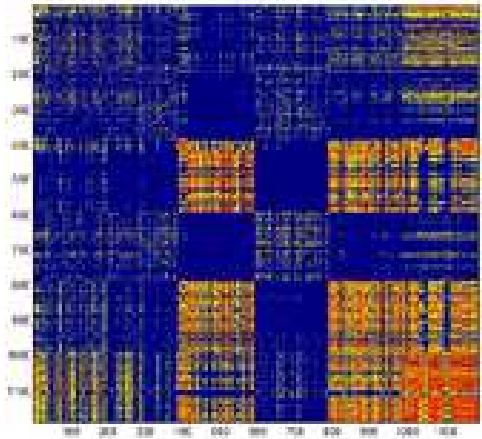}}\\
\end{tabular}
\caption{Correlation matrix of different types of force curves, described in the text. Color intensity from blue (low) to red (high) indicates the strength of correlation $C_{ij}$ between pairs of curves. Shown are correlations after matrix filtering and above a threshold $C_{ij} > 0.5$. }
\label{fig-bigmatrix}
\end{figure}

We created the correlation  matrix from all pairs of $N= 1188$ force curves, that were pre-processed in the  following way: contact part was removed, curves were corrected to the baseline, and only the part where interactions are expected was kept (first ~ 200nm). Curves for the correlation matrix were selected from four  different experiments with two different experimental setups.  The experiential setups differ in the  number of flexible linkers used to couple molecules to the surface and the cantilever: either both molecules were coupled to the cantilever/surface via PEG spacers or one was coupled to the gold surface
directly and the other to the cantilever via PEG spacer. In the correlation matrix (see Fig.\ \ref{fig-bigmatrix} and text below), from left-to-right, the first block (I) consists of 200 curves taken from measurements on RNA--Rev peptide
interaction (two linkers, velocity 581nm/s);  second block (II) consists of 188
curves taken from RNA--Rev peptide interaction in the presence of neomycin (two
linkers, velocity 581nm/s);  third block (III) contains 200 curves measured on RNA--Rev peptide interaction (one linker, velocity 2540nm/s);  fourth  block (IV) contains 200 curves  measured on ssDNA-ssDNA interaction (two linkers, velocity 218nm/s); fifth block (V) has  200 curves of RNA--Rev peptide interaction (one linker, velocity 1160nm/s),  and the last block (VI) consists of  200 curves of Rev peptide--Au interaction (one linker, velocity 1160nm/s).

\section{ Correlation Matrix and its Spectral Analysis}
In our dataset consisting of $N=1188$  force--distance curves, $\{f_i(x\})$, $i=1,2,3, \cdots N$, each curve is identified by a uniquely defined index $i$, thus representing a separate pulling event.
The elements of the correlation matrix $C_{ij}$ are  calculated as the Pearson's correlation coefficient between each pair  $(i,j)$ of curves as follows: 
\begin{equation}
 C_{ij}=\frac{\sum_{t_{k}}[f_{i}(x)-\langle f_{i}\rangle][f_{j}(x)-\langle f_{j}\rangle]}{\sigma_{i}\sigma_{j}} \ . \label{pearson}
\end{equation}
where the distance $x$ is given by a discrete set of measured values, and  $\sigma _i$, $\sigma _j$ stand for the standard deviation of the force signal $f_i(x)$ and $f_j(x)$. In Fig.\ \ref{fig-bigmatrix} we show a 3D color plot of the correlation matrix of all force curves, after filtering out  spurious correlations. For the filtering, we used the affinity transformation method \cite{eshel2004,tadic2009}, where the element is enhanced as $C_{ij}\to M_{ij}C_{ij}$ if the raws $i$ and $j$ correlate with the rest of the matrix in a similar way, yielding $M_{ij} >1$, and diminished otherwise. The meta-correlation factor $M_{ij}$ is computed as a Pearson's coefficient of the rearranged  elements $\{C_{ij}, C_{i1},...,C_{i-1j},C_{i+1j},\cdots ,C_{iN}\}$ and $\{C_{ji},C_{j1},...,C_{j-1i},C_{j+1i},\cdots ,C_{jN}\}$ without diagonal. 

For further discussion we notice that the  matrix can be represented by a network (mathematical graph), where each matrix index $i=1,2, \cdots N$ defines  network node and the matrix element $C_{ij}$---a link  between nodes $i$ and $j$. In our case the links are symmetrical $C_{ij}=C_{ji}$ by definition (\ref{pearson}). Notice that the matrix and network representations are formally equivalent. The network picture is suitable for visualization and topological interpretation. In particular, the matrix in Fig.\ \ref{fig-bigmatrix} makes a sparse network containing topological {\it modules}, i.e.,  groups of nodes with strong connections inside the group and sparse connections between them. As the Fig.\ \ref{fig-bigmatrix} shows, correlations between different sets of data (blocks I through VI) shown as off-diagonal block-matrices can be as strong as correlations inside the same data set (diagonal blocks). This suggests that the network modules may contain curves from different data blocks! In the following we apply the eigenvalue spectral methods to identify these topological modules.

Here we perform spectral analysis  of the normalized Laplacian operator $\mathbf{L}$ related to the filtered correlation  $\mathbf{C}$ in Fig.\ \ref{fig-bigmatrix}, or more precisely, its binary form with the elements:  $A_{ij} =1$ whenever  $C_{ij} >C_0$, or $A_{ij}=0$ otherwise. The matrix elements of the Laplacian are given by \cite{mitrovic2008b,samukhin2007}
\begin{equation}
L_{ij}=\delta_{ij}-\frac{A_{ij}}{\sqrt{q_{i}q_{j}}} \ , \label{lap3}
\end{equation}
where $q_i$, $q_j$ are the number of links at nodes $i$ and $j$, respectively.
Although the same conclusions can be reached using the adjacency matrix $A_{ij}$ directly, the analysis of the Laplacian (\ref{lap3}) is more convenient since its eigenvalue spectrum  is limited in the range $\lambda _i \in [0,2]$. Moreover, its  {\it eigenvectors belonging to few lowest nonzero eigenvalues tend to localize along the network modules} \cite{donetti2004,mitrovic2008b}. This is a direct consequence of the orthogonality to the eigenvector belonging to zero eigenvalue of the Laplacian (or the largest eigenvalue of the adjacency matrix) which has all positive components (Perron-Frobenius theorem \cite{agth-graphtheory}).
Precisely, the localization  means that, among $N$ components of the eigenvector, the nonzero (positive/negative) components  have indexes which coincide with nodes in a network module (detailed analysis of spectra in modular networks is given in \cite{mitrovic2008b}).  
We use this property of the eigenvectors to identify nodes in different modules.

\begin{figure}
\centering
\begin{tabular}{cc}
\resizebox{16pc}{!}{\includegraphics{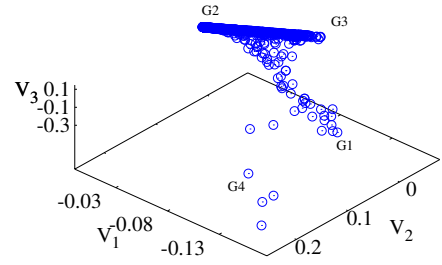}}\\
\end{tabular}
\caption{Scatterplot of the eigenvectors $V_1,V_2,V_3$ of three lowest eigenvalues of the Laplacian operator (\ref{lap3}) related to the full correlation matrix in Fig.\ \ref{fig-bigmatrix}. Four branches are visible in this projection, marked as $G_1,G_2,G_3,G_4$.}
\label{fig-scatterbig}
\end{figure}
\begin{figure}[!]
\centering
\begin{tabular}{cc}
\resizebox{18pc}{!}{\includegraphics{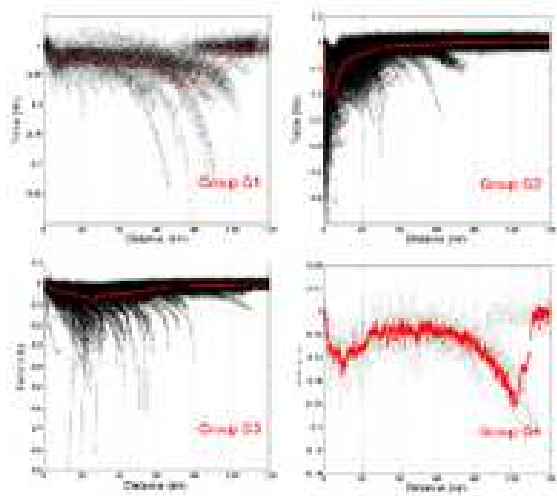}}\\
\end{tabular}
\caption{Overlap plot of all curves belonging to the groups $G_1,G_2,G_3,G_4$ identified from the scatterplot in Fig.\ \ref{fig-scatterbig}.}
\label{fig-overlaybig}
\end{figure}

\section{Identification of modules: Force-curves grouping}
When a modular structure occurs, the localization of eigenvectors belonging to smallest non-zero eigenvalues is manifested in a {\it branched} pattern of the scatter-plot in the space of these eigenvectors \cite{mitrovic2008b}. In Fig.\ \ref{fig-scatterbig} we show the scatterplot of the eigenvectors $(V_1,V_2,V_3)$ belonging for three smallest non-zero eigenvalues of the Laplacian (\ref{lap3}), related to the correlation matrix in Fig.\ \ref{fig-bigmatrix}. In the scatterplot, each point carries one index, thus indicating one network node, that is a force curve index in our original dataset. The plot in Fig.\ \ref{fig-scatterbig} shows four branches, marked by $G_1,G_2,G_3,G_4$, thus four groups of curves can be identified.  The most representative curves in each group are those at the tips of branches with most distinct curves sitting  at the opposite ends of the branches $G_2$ and $G_3$,  whereas, the differences gradually diminishes closer to the center of the plot.  By matching the indexes with curves in the original dataset, we identify representative curves at the tips of four branches (Table \ 1.). Overlay of all curves in the $G_1 \cdots G_4$ groups is shown in Fig.\ \ref{fig-overlaybig}.

\begin{table}
\caption{(upper part)Representative curves appearing at tips of four branches $G_1,G_2,G_3,G_4$ in the scatterplot in Fig.\ \ref{fig-scatterbig} and their distribution over original blocks od data I through VI. (Middle and bottom) Further splitting of groups $G_2$ and $G_3$. }
\label{tab.1}
\begin{center}
\begin{tabular}{|c|cccccc|}
\hline
block  &     I &          II  &         III &           IV  &           V   &       VI\\\hline\hline
 G1&       5   &         9   &         0  &          16 &           0 &          0\\
G2   &      51&          28 &         177 &        0 &            181 &       148\\
G3  &       60 &         68&           0 &         135 &           1   &        7\\
G4 &       4 &             0   &        1  &          0   &           0  &          0\\\hline
G2-g1  &  5  &   1 &      3 &     0 &      35 &     22\\
G2-g2  &  0 &    0   &   45   &  0   &    7   &     0\\
G2-g3&20   & 10  &  5  &    0  &     3 &       21\\
\hline
G3-g1 & 5& 6& 0& 24& 1& 0\\
G3-g2 & 24& 9& 0& 24& 0& 5 \\
G3-g3 & 18& 15& 0& 28& 0&1\\
\hline\hline
\end{tabular}
\end{center}
\end{table}

\begin{figure}[!]
\centering
\begin{tabular}{cc}
\resizebox{16pc}{!}{\includegraphics{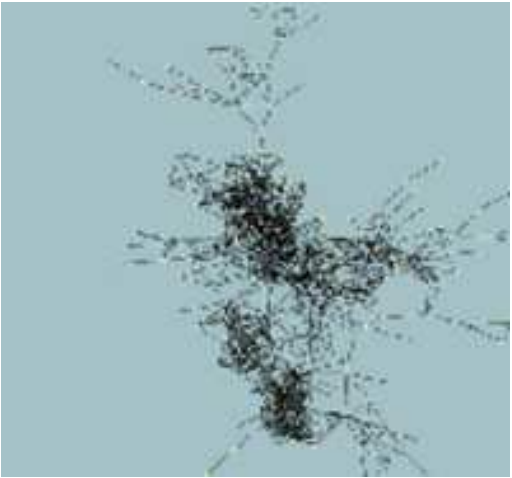}}\\
\end{tabular}
\caption{Example of the correlation network constructed from the force curves in group $G_2$ from Fig.\ \ref{fig-scatterbig}, exhibiting modular structure. Shown are only links above the threshold $C_0=0.91$.}
\label{fig-netG2}
\end{figure}

Each of the groups contains number of curves from one or several blocks of the original datasets. Recalling the nature of the data in different blocks, we see that the module $G_2$ contains curves obtained predominantly on experimental setup with one PEG  spacer (blocks III, V, and VI) and a fraction of data with two spacers (from blocks I and II), while excluding altogether the DNA-DNA  interactions (block IV). On the other hand, the module $G_3$ contains data from DNA-DNA  interactions  and a number of curves from Rev-RNA and blocked Rev-RNA with neomycin, all data obtained using two spacers.  $G_1$ has similar composition although it appears as a separate module, similar as the small group $G_4$ (see visual differences in Fig.\ \ref{fig-overlaybig}).

\begin{figure}
\centering
\begin{tabular}{cc}
\resizebox{14pc}{!}{\includegraphics{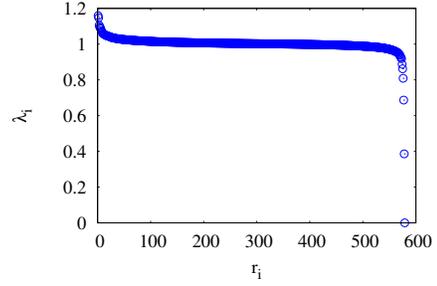}}\\
\resizebox{16pc}{!}{\includegraphics{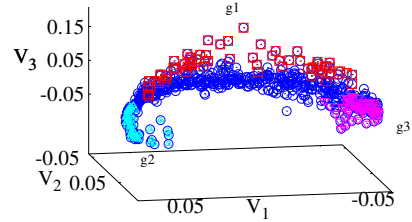}}\\
\end{tabular}
\caption{For the branch $G_2$ of Fig.\ \ref{fig-scatterbig}: (top) the eigenvalue spectrum and (bottom) the scatterplot of three eigenvectors belonging to lowest nonzero eigenvalues. Three subgroups are identified marked by $g1,g2,g3$, see text for details.}
\label{fig-scatterG2}
\end{figure}

We further analyze the group of curves in  $G_2$ applying the same approach on the now reduced correlation matrix. The complete group consists of $N_2=597$ curves. The correlation network of these curves, shown in Fig.\ \ref{fig-netG2}, exhibits modular structure, suggesting that smaller subgroups of the group $G_2$ can be identified. The eigenvalues of the Laplacian related to this correlation matrix are shown in the ranking order in Fig.\ \ref{fig-scatterG2}, top panel. Compatible with the network modularity in Fig.\ \ref{fig-netG2} is the appearance of three eigenvalues in the gap between  $\lambda =0$ and the main part of the spectrum.  The corresponding scatterplot in the space of three eigenvectors belonging to these eigenvalues is also shown in Fig.\ \ref{fig-scatterG2}, bottom panel. In this case similarity between points, forming a half-helmet in $(V_1,V_2,V_3)$ space, is stronger compared to Fig.\ \ref{fig-scatterbig}, however, groups can be identified by the end points of the half-ring in the horizontal plane and points with largest vertical distance, marked as $g_1,g_2,g_3$ and different symbols (colors) in Fig.\ \ref{fig-scatterG2}.

\begin{figure}
\centering
\begin{tabular}{cc}
\resizebox{18pc}{!}{\includegraphics{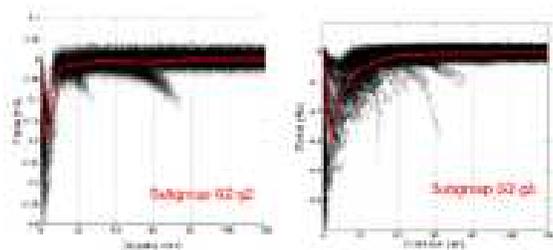}}\\
\end{tabular}
\caption{Overlay of the selected force curves of low- and high-velocity associated to ends of the scatterplot ring in Fig.\ \ref{fig-scatterG2}.}
\label{fig-overlayG2g}
\end{figure}

The identity of these curves with respect to the original datasets is also indicated in Table\ 1, middle part. As mentioned above, curves in large group $G_2$ are on Rev peptide binding, whereas, their subgroups  appear to be from different pulling velocities: $G_2-g1$ consists mostly of curves in blocks V and VI, measurements at velocity of 1160 nm/sec, while the groups $G_2-g2$ and $G_2-g3$ at two ends of the semi-ring contains curves measured at highest (2540 nm/sec) and lowest (581 nm/sec) velocity, respectively. In the experiment, measurements at different velocities and correct assignments of the curves are important for the extrapolation of bonding parameters to zero force values. Pulling velocity affects both force loading rate $\dot{F}(F)$ as well as the average disruption force $\langle F\rangle$. 
Within our methods selected groups of high- and low-velocity  curves, $G_2-g2$ and $G_2-g3$, respectively, are shown in Fig.\ \ref{fig-overlayG2g}. Similar analysis of the data contained in the group $G_3$ (not shown) leads to the curves identified in the lower part of the Table \ 1. Here data contain three different interactions (DNA-DNA, Rev-RNA, and Rev-RNA-with-neomycin), all measured in the setup with two spacers and low velocity. We find much larger mixing between the groups, indicating either increased number of aspecific binding or a dominant role of spacers.

\begin{figure}
\centering
\begin{tabular}{cc}
\resizebox{16pc}{!}{\includegraphics{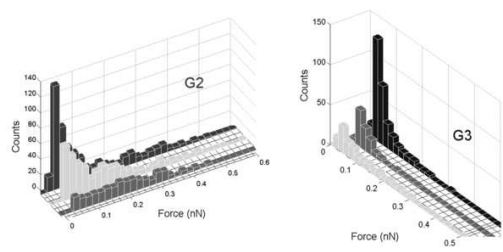}}&
\end{tabular}
\caption{Histograms of rupture forces from selected curves in blocks 
VI, V, III  from group $G_2$  (left) and from blocks I, II, IV from group $G_3$ (right). ( Cf. Table\ 1. and data description.)}
\label{fig-histogramsG2}
\end{figure}

Finally, in Fig.\ \ref{fig-histogramsG2} we show histograms of the rupture forces made upon curves in groups $G_2$ and $G_3$ selected by our methodology. Detailed analysis of Rev--RNA bonding parameters is given in \cite{Jelena-experimental2009}. 

\section{Conclusions}
We have shown that stochastic signals of different molecular systems (peptide--RNA, DNA--DNA) measured in the dynamic force spectroscopy experiments can be effectively selected according to their similarities. Our methodology uses the signal's relevant correlation matrix mapped onto a mathematical graph. Then using the spectral analysis of these graphs the groups of similar curves are detected, which appear as topological modules on them. Although the method is essentially statistical, we have shown that strong regularities in groupings of the force curves occur (and can be effectively used for the signal evaluation), based on the pulling speeds and experimental setup and even type of the interaction measured. Note that for the demonstration purposes in this work we used raw data with minimal pre-processing. Further pre-processing, e.g., with cuve-fitting \cite{fuhrmann2008}, could be incorporated and increase selectivity in our method. Improved efficiency in the evaluation of force curves by this approach is also expected   in the case of larger molecules or molecular complexes and the situations where distinction between many binding sites is applicable.

\acknowledgments
Work supported by the FP6 project Functional and Structural Genomics on Viral RNA: FSG-V-RNA, the FP7 project CYBEREMOTIONS, and the  Nanotechnology programme of the Ministry of Economic Affairs, NanoNed (Netherlands) and the  national program P1-0044 (Slovenia).  J.\v Z. also thanks for the hospitality during her stay at the J. Stefan Institute, Ljubljana.


\begin{thebibliography}{10}

\bibitem{barkai2008}
E.~{Barkai}, F.~{Brown}, M.~{Orrit}, and H.~{Yang}, {\em {Theory and evaluation
  of single-molecule signals}}.
\newblock Worls Scientific, Singapur, 2008.

\bibitem{evansritchie1997}
E.~{Evans} and K.~{Ritchie}, ``Dynamic strength of molecular adhesion bonds,''
  {\em Biophysical Journal}, vol.~72, pp.~1541--1555, 1997.

\bibitem{strunz1999}
T.~{Strunz}, K.~{Oroszlan}, R.~{Sch\"afer}, and H.~{G\"untherodt}, ``Dynamic
  force spectroscopy of single dna molecule,'' {\em PNAS}, vol.~96, no.~41,
  pp.~11277--11282, 1999.

\bibitem{DSFreview2000}
J.~{Zlatanova}, S.~{Lindsay}, and S.~{Leuba} {\em Prog. Biophys. Mol. Biol.},
  vol.~74, p.~37, 2000.

\bibitem{koch2003}
S.~{Koch} and M.~{Wang}, ``Dynamic force spectroscopy of protein-dna
  interactions by unzipping dna,'' {\em Phys. Rev. Lett.}, vol.~91, p.~028103,
  Jul 2003.

\bibitem{riet2007}
J.~{te Riet}, A.~{Zimmerman}, A.~{Cambi}, B.~{Joosten}, S.~{Speller},
  R.~{Torensma}, F.~{van Leeuwen}, C.~{Figdor}, and F.~{de Lange}, ``Distinct
  kinetic and mechanical properties govern alcam-mediated interactions as shown
  by single-molecule force spectroscopy,'' {\em Journal of Cell Science},
  vol.~120, pp.~3965--3976, 2007.

\bibitem{sulchek2005}
T.~{Sulchek}, R.~{Friddle}, K.~{Langry}, E.~{Lau}, H.~{Albrecht}, V.~{Ratto},
  S.~{DeNardo}, M.~{Colvin}, and A.~{Noy}, ``Dynamic force spectroscopy of
  parallel individual mucin1-antibody bonds,'' {\em PNAS}, vol.~102,
  pp.~16638--16643, 2005.

\bibitem{RNA2009}
A.~{Fuhrmann}, J.~{Schoening}, D.~{Anselmetti}, D.~{Staiger}, and R.~{Ros},
  ``Quantitative analysis of single-molecule rna-protein interaction,'' {\em
  Biophysical Journal}, {\bf 96}, 5030, 2009.

\bibitem{raible2006}
M.~{Raible}, M.~{Evstignev}, F.~{Bartels}, R.~{Eckel}, M.~{Nguyen-Duong},
  R.~{Merkel}, R.~{Ros}, D.~{Anselmetti}, and P.~{Reimann}, ``Theoretical
  analysis of single-molecule force spectroscopy experiments: Heterogeneity of
  chemical bonds,'' {\em Biophysical Journal}, vol.~90, p.~3851, 2006.

\bibitem{dudko2006}
O.~{Dudko}, G.~Hummer, and A.~Szabo, ``Intrinsic rates and activation free
  energies from single-molecule pulling experiments,'' {\em Phys. Rev. Lett.},
  vol.~96, p.~108101, 2006.

\bibitem{dudko2008}
O.~{Dudko}, G.~Hummer, and A.~Szabo, ``Theory, analysis, and interpretation of
  single-molecule force spectroscopy experiments,'' {\em PNAS}, vol.~105,
  pp.~15755--15760, 2008.

\bibitem{fuhrmann2008}
A.~{Fuhrmann}, D.~{Anselmetti}, and R.~{Ros}, ``Refined procedure of evaluating
  experimental single-molecule force spectroscopy data,'' {\em Phys. Rev. E},
  {\bf 77}, 031912, 2006.

\bibitem{marsico2006}
A.~{Marsico}, D.~{Labudde}, T.~{Sapra}, D.~{Muller}, and M.~{Schroeder}, ``A
  novel pattern recognition algorithm to classify membrane protein unfolding
  pathways with high-throughput single-molecule force spectroscopy,'' {\em
  Bioinformatics}, vol.~23, no.~2, pp.~e231--e236, 2005.

\bibitem{boccaletti}
S.~Boccaletti, V.~Latora, Y.~Moreno, M.~Chavez, and D.~U. Hwang, ``Complex
  networks: Structure and dynamics,'' {\em Physics Reports}, vol.~424, p.~175,
  2006.

\bibitem{donetti2004}
L.~{Donetti} and M.~A. {Mu{\~n}oz}, ``{Detecting network communities: a new
  systematic and efficient algorithm},'' {\em Journal of Statistical Mechanics:
  Theory \& Experiment}, {\bf 10}, 2004.

\bibitem{mitrovic2008b}
M.~{Mitrovi\'c} and B.~{Tadi\'c}, ``Spectral and dynamical properties in
  classes of sparse networks with mesoscopic inhomogeneities,'' {\em Physical
  Review E}, vol.~80, no.~1, 2009.

\bibitem{benjacob2006}
I.~{Baruchi}, D.~{Grossman}, V.~{Volman}, M.~{Shein}, J.~{Hunter}, V.~{Towle},
  and E.~{Ben-Jacob}, ``Functional holography analysis: Simplifying the
  complexity of dynamical networks,'' {\em Chaos}, vol.~16, p.~015112, 2006.

\bibitem{mantegna1999}
R.~N. {Mantegna}, ``{Hierarchical structure in financial markets},'' {\em
  European Physical Journal B}, vol.~11, p.~193, 1999.

\bibitem{zivkovic2006}
J.~{{\v Z}ivkovi{\'c}}, B.~{Tadi{\'c}}, N.~{Wick}, and S.~{Thurner},
  ``{Statistical indicators of collective behavior and functional clusters in
  gene networks of yeast},'' {\em European Physical Journal B}, vol.~50,
  pp.~255--258, 2006.

\bibitem{zivkovic2009}
J.~{\v Z}ivkovi{\'c}, M.~{Mitrovi\'c}, and B.~{Tadi\'c}, ``{Correlation
  Patterns in Gene Expression along the Cell Cycle of Yeast},'' {\em Springer
  Series: Studies in Computational Intelligence}, vol.~207, pp.~23--34, 2009.

\bibitem{zemanova2000}
L.~Zemanova, C.~Zhou, and J.~Kurths, ``Structural and functional clusters of
  complex brain networks,'' {\em Physica D: Nonlinear Phenomena}, vol.~224,
  pp.~202--212, 2006.

\bibitem{eshel2004}
I.~{Baruchi} and E.~{Ben-Jacob}, ``{Functional holography of recorded neuronal
  networks activity},'' {\em Neuroinformatics}, vol.~2, no.~3, pp.~333--351,
  2004.

\bibitem{book_kurths2008}
R.~{Graben}, C.~{Zhou}, M.~{Thiel}, and J.~{Kurths}, {\em {Lectures in
  Supercomputational Neuroscience: Dynamics in Complex Brain Networks
  (Understanding Complex Systems}}.
\newblock Springer-Verlafg, Berlin Heidelberg, 2008.

\bibitem{shlomo2009}
K.~{Yamasaki}, A.~{Gozolchiani}, and S.~{Havlin}, ``Climat networks based on
  phase synchronization analysis track el-nino,'' {\em Prog. Theor. Phys.
  Suppl.}, vol.~179, p.~178, 2009.

\bibitem{tadic2009}
B.~{Tadi\'c} and M.~{Mitrovi\'c}, ``Jamming and correlation patterns in traffic
  of information on sparse modular networks,'' {\em European Physical Journal
  B}, 2009.

\bibitem{Jelena-experimental2009}
J.~{\v Z}ivkovi{\'c}, L.~{Jansen}, F.~{Alvarado}, H.~{Heus}, and S.~{Speller},
  ``{Force spectroscopy on RRE-Rev complex from HIV1},'' {\em in preparation}, 2009.

\bibitem{samukhin2007}
A.~N. {Samukhin}, S.~N. {Dorogovtsev}, and J.~F.~F. {Mendes}, ``{Laplacian
  spectra of, and random walks on, complex networks: Are scale-free
  architectures really important?},'' {\em Physical Review E}, vol.~77, no.~3,
  pp.~036115--+, 2008.

\bibitem{agth-graphtheory}
C.~{Godsil} and G.~{Royle}, {\em {Algebraic Graph Theory}}.
\newblock Springer-Verlafg, Berlin Heidelberg, 2001.

\end{thebibliography}
\end{document}